\documentclass[10pt]{article}
\usepackage{fullpage}
\usepackage[english]{babel}
\usepackage[utf8]{inputenc}
\usepackage{amsmath,amssymb,mathtools,xy,lscape} 
\usepackage{amsmath, amssymb, bm}
\usepackage{graphicx}
\usepackage{subcaption}
\usepackage[colorinlistoftodos]{todonotes}
\usepackage{indentfirst}
\xyoption{all}    

\newcommand{\hiddenpower}[2] { \ifnum \numexpr#2=1 #1 \else #1^#2 \fi }
\numberwithin{equation}{section}

\usepackage{xparse}
\ExplSyntaxOn
\newcounter{diff_order}
\newcounter{diff_power}

\newcommand{\rawdiff}[3]
{
	\setcounter{diff_order}{0}
	\clist_map_inline:nn{#3}{\stepcounter{diff_order}}
	
	\frac{\hiddenpower{#1}{\thediff_order} #2}
	{
		\def\old_var{DefaultValue}
		\setcounter{diff_power}{0}
		
		\clist_map_inline:nn{#3}
		{
			\def\new_var{##1}
			\ifnum \thediff_power=0
				\stepcounter{diff_power}
			\else
				\tl_if_eq:NNTF \new_var \old_var
				{\stepcounter{diff_power}}
				{
					#1 \hiddenpower{\old_var}{\thediff_power}
					\setcounter{diff_power}{1}
				}
			\fi

			\def\old_var{##1}
		}
		
		#1 \hiddenpower{\old_var}{\thediff_power}
	}
}

\ExplSyntaxOff

\newcommand{\lb}{\left(}
\newcommand{\rb}{\right)}

\renewcommand{\sin}[2][1]{\hiddenpower{\text{sin}}{#1} \lb #2 \rb}

\renewcommand{\cos}[2][1]{\hiddenpower{\text{cos}}{#1} \lb #2 \rb}

\renewcommand{\ln}[1]{\text{ln} \lb #1 \rb}

\begin{document}

\begin{center}
\strut\hfill



\noindent {{\bf{TIME EVOLUTION IN QUANTUM SYSTEMS AND STOCHASTICS}}}\\
\vskip 0.3in

\noindent {{\footnotesize ANASTASIA DOIKOU\footnote{{\tt This is based on a talk given by AD in ``Quantum Theory and Symmetries XI'', 
July 2019, Montreal, Canada.}}, SIMON J.A. MALHAM AND ANKE WIESE }}
\vskip 0.4in

\noindent {\footnotesize School of Mathematical and Computer Sciences, Department of Mathematics,\\ 
Heriot-Watt University, Edinburgh EH14 4AS, United Kingdom}

\vskip 0.1in
\noindent {\footnotesize {\tt E-mail: a.doikou@hw.ac.uk,\  s.j.a.malham@hw.ac.uk \ a.wiese@hw.ac.uk}}\\

\vskip 0.60in

\end{center}

\begin{abstract}
\noindent The time evolution problem for non-self adjoint second order differential operators is studied 
by means of  the path integral formulation.  
Explicit computation of the path integral via the use of certain underlying stochastic differential equations, 
which naturally emerge when computing the path integral,  
leads to a universal expression for the associated measure regardless of the form of the differential operators. 
The discrete non-linear hierarchy (DNLS) is then considered and the corresponding hierarchy 
of solvable, in principle, SDEs is extracted. The first couple members of the hierarchy correspond 
to the discrete stochastic transport and heat equations. 
The discrete stochastic Burgers equation is also obtained through 
the analogue of the Cole-Hopf transformation. 
The continuum limit is also discussed.
\end{abstract}

\section{Introduction}

\noindent One of our main aims here is the solution of the time evolution problem associated to 
non self-adjoint operators using the path integral formulation.
We consider the general second order differential operator $ \hat L_0$, 
and the associated time evolution problem:  
\begin{eqnarray}
&&-\partial_t f({\mathrm x},t) = \hat L f({\mathrm x},t)= \Big (\hat L_0 + 
u({\mathrm x})\Big ) f({\mathrm x},t), \label{gen0}\\
&& \hat L_0 = {1 \over 2}\sum_{i, j=1}^M g_{ij}({\mathrm x}) {\partial^2
\over \partial{{\mathrm x}_i} \partial{{\mathrm x}_j}} + 
\sum_{j=1}^M b_j({\mathrm x}){\partial \over \partial{{\mathrm x}_j} } , 
~~~~~~g({\mathrm x}) = \sigma({\mathrm x})\sigma^T({\mathrm x})
 \label{general}
\end{eqnarray}
where the diffusion matrix $g({\mathrm x})$ and the matrix $\sigma({\mathrm x})$ are in general dynamical 
(depending on the fields ${\mathrm x}_j$) $M \times M$ matrices, 
while ${\mathrm x}$ and the drift $b({\mathrm x})$ are $M$ vector fields with components ${\mathrm x}_j$, $b_j$ 
respectively, and $^T$ denotes usual transposition. 
The operator $\hat L_0$ is not in general self-adjoint (Hermitian), therefore 
we also introduce the adjoint operator defined for any suitable function $f({\mathrm x}, t)$ as
\begin{eqnarray}
\hat L_0^{\dag}f({\mathrm x}, t) ={1 \over 2}\sum_{i, j=1}^M  {\partial^2\over \partial{{\mathrm x}_i} 
\partial{{\mathrm x}_j}}\Big (g_{ij}({\mathrm x}) f({\mathrm x}, t)\Big ) - 
\sum_{j=1}^M {\partial \over \partial{{\mathrm x}_j}}\Big (b_j({\mathrm x}) f({\mathrm x}, t)\Big ).\label{adjoint}
\end{eqnarray}
Then, two distinct time evolution equations emerge.
\\
\noindent 1. The Fokker-Plank equation:\\
\begin{equation}
\partial_{t_1} f({\mathrm x},t_1) = \hat L_0^{\dag} f({\mathrm x},t_1) \label{fokker} 
\end{equation}
$ t_1\geq t_2$ , with known initial condition $f({\mathrm x}, t_2)=f_{0}({\mathrm x})$. 
\\
\noindent 2. The  Kolmogorov backward equation:\\ 
\begin{equation}
-\partial_{t_2} f({\mathrm x},t_2) = \hat L_0 f({\mathrm x},t_2) \label{back} 
\end{equation}
 $t_2 \leq t_1$, with known final condition $f(x, t_1)=f_{f}(x)$.

To simplify the complicated situation of non-constant diffusion coefficients  we 
employ a local change of frame, which 
reduces the $\hat L$ operator to the simpler form with constant diffusion coefficients,
and an effective drift. 
We then compute the generic path integral 
by requiring that the fields involved satisfy discrete time analogues of stochastic differential equations (SDEs). 
These equations naturally emerge when computing the path integral via the time discretization scheme. 
This  leads to the computation of the measure, which turns out to be an infinite product of Gaussians. 

Our second objective is to explore links between SDEs, and quantum integrable systems.
To illustrate these associations we discuss a typical exactly solvable discrete quantum system, 
the discrete non-linear Schr\"odinger hierarchy. We express the quantum integrals of motion as second order 
differential operators after a suitable rescaling of the fields and we then extract a hierarchy of associated SDEs,
which can be in principle solved by means of suitable integrator factors. The first two non-trivial members of the hierarchy 
correspond to the discrete stochastic transport and heat equations.
The discrete stochastic Burgers equation is also obtained from the discrete stochastic heat equation through 
the analogue of the Cole-Hopf transformation  (see also relevant \cite{Corwin}).
More details on the derivation of the reported results can be found in \cite{DoikouMalhamWiese}.

\section{Time evolution and the Feynman-Kac formula}

\noindent  Before we compute the solution of the time evolution problem via the path integral formulation
 we shall implement the quantum canonical transform, that turns the dynamical diffusion matrix in (\ref{general}) 
into identity at the level of the PDEs. This result will be then used  for the explicit computation of the general
path integral, and the derivation of the Feynman-Kac formula \cite{DoikouMalhamWiese}.

\subsection{The quantum canonical transformation}

\noindent We will show in what follows that the general $\hat L$ operator can be brought into the less involved form:
\begin{equation}
\hat L = {1\over 2} \sum_{j=1}^M {\partial^2 \over \partial{{\mathrm y}_j}^2} +
\sum_{j=1}^M  \tilde b_j({\mathrm y}) {\partial \over \partial{{\mathrm y}_j}}+u({\mathrm y}) \label{constant}
\end{equation}
with an induced drift $\tilde b({\mathrm y})$. This can be achieved via a simple change of the parameters ${\mathrm x}_j$, 
which geometrically is nothing but a change of frame. 
Indeed, let us introduce a new set of parameters ${\mathrm y}_j$ such that \cite{DoikouMalhamWiese}:
\begin{equation}
 d{\mathrm y}_i = \sum_j \sigma^{-1}_{ij}({\mathrm x})\ d{\mathrm x}_j, ~~~~\det \sigma \neq 0, \label{frame}
\end{equation}
then $\hat L$ can be expressed in the form (\ref{constant}), and the induced drift components are given as
\begin{equation}
\tilde b_k({\mathrm y}) = \sum_j \sigma_{kj}^{-1}({\mathrm y}) b_j({\mathrm y}) +{ 1\over 2}
 \sum_{j,l} \sigma_{jl}({\mathrm y}) \partial_{{\mathrm y}_l} \sigma_{kj}^{-1}({\mathrm y}). \label{drifta}
\end{equation}
Bearing also in mind that $\sum_j \sigma_{jl} \sigma_{kj}^{-1} = \delta_{kl}$,
we can write in the compact vector/matrix notation:
\begin{equation}
\tilde b({\mathrm y}) = \sigma^{-1}({\mathrm y})\Big(b({\mathrm y}) -{ 1\over 2}  (\nabla_{{\mathrm y}} 
\sigma^T({\mathrm y}))^T\Big ), ~~~~~~
\nabla_{\mathrm y} = \Big (\partial_{{\mathrm y}_1}, \ldots, \partial_{{\mathrm y}_M} \Big ) \label{drift1}
\end{equation}
where one first solves for ${\mathrm x} ={\mathrm x}({\mathrm y})$ via (\ref{frame}). The transformation discussed above corresponds to a generalization 
of the so called Lamperti transform at the level of SDEs (we refer the interested reader to \cite{DoikouMalhamWiese} and references therein).

\subsection{The path integral: Feynmann-Kac formula}

\noindent We are now in the position to solve the time evolution problem for the considerably simpler operator (\ref{constant}).
Our starting point is the time evolution equation (\ref{adjoint}), (\ref{fokker}),  (\ref{constant}):
\begin{equation}
\partial_t f({\mathrm y} ,t) = \hat L^{\dag}  f({\mathrm y}, t), \label{evolution} \nonumber
\end{equation}
we then explicitly compute the propagator $K({\mathrm y}_f, {\mathrm y}_i| t, t')$:
\begin{eqnarray}
f({\mathrm y} ,t) &=& \int  \prod_{j=1}^M d y_j'\ K({\mathrm y}, {\mathrm y}'| t, t')f({\mathrm y}' ,t') \label{kernel0}\\
&=& \int \prod_{n=1}^N \prod_{j=1}^M d y_{jn}\ 
\prod_{n=1}^N K({\mathrm y}_{n+1}, {\mathrm y}_n| t_{n+1}, t_n)f({\mathrm y}_1 ,t_1). \label{kernel}
\end{eqnarray}
We employ the standard time discretization scheme as shown above, (see also for instance \cite{Simon}), we insert the unit $N$ times, 
(${1\over 2 \pi} \int d{\mathrm y}_{jn}\  dp_{jn}\ e^{ {\mathrm i} p_{jn}({\mathrm y}_{jn}-a)}=1$), for each component ${\mathrm y}_j$, and 
we perform the Gaussian integrals with respect to each $p_{jn}$ parameter. We then conclude that the path integral can be expressed as
\begin{eqnarray}
&& K({\mathrm y}_f, {\mathrm y}_i| t, t') = \int d{\bf q}\ \exp\Big [- \sum_j \sum_{n}{\big (\Delta {\mathrm y}_{jn} -\delta 
\tilde b_{jn}({\mathrm y})\big)^2 \over 2\delta} +
\delta\sum_n  u_n({\mathrm y})\Big ] \label{exp}\\
&&  d{\bf q}  ={1 \over (2 \pi \delta )^{NM\over 2}} \prod_{n=2}^N \prod_{j=1}^M  d{\mathrm y}_{jn} \label{before}
\end{eqnarray}
where $f_{n} = f_n({\mathrm y}_n)$ and $\Delta {\mathrm y}_{jn} = {\mathrm y}_{jn+1} - {\mathrm y}_{jn}$.
where $\delta= t_{n+1}- t_{n}$ and with boundary conditions: ${\mathrm y}_f = {\mathrm y}_{N+1},  ~~
{\mathrm y}_i = {\mathrm y}_{1}, ~~t_i =t'=0$
($t'$ will be dropped henceforth for brevity), $~t_f=t$.

We recall expression (\ref{exp}) and we make the fundamental assumption \cite{DoikouMalhamWiese}:
\begin{equation}
\Delta {\mathrm y}_n - \delta \tilde b_n({\mathrm y}) =  \Delta {\mathrm w}_n \label{DSDE0}
\end{equation}
assuming also that ${\mathrm w}_{nj}$ are Brownian paths (see for instance \cite{Prevot} on Wiener processes),
 i.e. (\ref{DSDE0}) is the discrete time analogue of an SDE.
After a change of the volume element in (\ref{exp}), subject to (\ref{DSDE0}), we conclude 
(see \cite{DoikouMalhamWiese} for the detailed computation):
\begin{equation}
 K({\mathrm y}_f, {\mathrm y}_i| t)=\int  d {\bf M}\ e^{\int_{0}^t u({\mathrm y}_s) ds}, \label{B}
\end{equation}
\begin{equation} 
d {\bf M} = \lim_{\delta \to 0} \lim_{ N \to \infty}\ {1 \over (2 \pi \delta )^{NM\over 2}}\
\exp\Big [- {1\over 2 \delta} \sum_{n=1}^N \Delta {\mathrm w}_n^T \Delta {\mathrm w}_n \Big  ]\
 \prod_{n=2}^N \prod_{j=1}^M d {\mathrm w}_{jn}. \label{dmeasure}
\end{equation}
We may now evaluate the measure:  in the continuum time limit (\ref{dmeasure}), we consider the Fourier representation 
on $[0,\  t]$ for ${\mathrm w}_s$, i.e. Wiener's representation of the Brownian path \cite{Prevot}:
\begin{equation}
{\mathrm w}_s ={ {\mathrm f}_0\over \sqrt{t} } s + \sqrt{2 \over t} \sum_{k >0} { {\mathrm f}_k\over \omega_k} \sin{ \omega_ks},
 ~~~~\omega_k = {2\pi k \over t}. \label{random}
\end{equation}
${\mathrm f}_0 = {{\mathrm w}_t\over\sqrt{t}  }$ and ${\mathrm f}_k,\ k\in \{0,\ 1,  \ldots\}$ 
are $M$ vectors with components ${\mathrm f}_{kj},\ j \in \{1,\  2,  \ldots,  M\}$  being  standard normal variables.  
We are interested in the computation of the measure in the continuum limit 
$N\to \infty,\ \delta \to 0$, and we also recall the following boundary conditions: ${\mathrm w}(s=0) = 0,\ {\mathrm w}(s=t) ={\mathrm w}_t$, then
\begin{eqnarray}
&& d {\bf M} = {e^{-{1\over 2t}  {\mathrm w}^T_{t}{\mathrm w}_{t}} \over (2\pi t)^{{M\over 2}}}\  d{\bf M}_0\nonumber\\
&& d {\bf M}_0 =  \prod_{k \geq 1} \prod_{j=1}^M{ d{\mathrm f}_{kj} \over \sqrt {2\pi }}\  
\exp[-{1\over 2} \sum_{k\geq1}\sum_j {\mathrm f}_{kj}^2]. \label{measure1}
\end{eqnarray}
The measure naturally is expressed as an infinite product of Gaussians regardless of the specific 
forms of the diffusion coefficients and the drift.

Having computed the propagator explicitly (\ref{B}) we conclude that equation (\ref{kernel}) 
can be then expressed as
\begin{equation}
f({\mathrm x}_f, t_f) =  \int d {\bf M}\ e^{\int_0^t u({\mathrm x}_s)ds} f_0({\mathrm x}_0), ~~~~~
f_0({\mathrm x}_0) = f({\mathrm x}_0, t_0) \nonumber
\end{equation}
which  is precisely  the Feynman-Kac formula, and describes the time evolution of a given initial profile 
$f_0({\mathrm x}_0)$ to $f({\mathrm x}_f, t_f) $ a solution of the Fokker-Planck equation.
One could have started from the Kolmogorov backward equation and computed the path integral backwards in time:
\begin{equation}
f({\mathrm x}_0, t_0) =  \int d {\bf M}\ e^{\int_0^t u({\mathrm x}_s)ds} f_f({\mathrm x}_f), ~~~~~
f_f({\mathrm x}_f) = f({\mathrm x}_f, t_f). \nonumber
\end{equation}
In this case the Feynman-Kac formula describes the reversed time evolution of a given final state $f_f({\mathrm x}_f)$,  
to a previous state $f({\mathrm x}_0, t_0) $  a solution of the Kolmogorov backward equation.

One of the main aims is the computation of expectation values:
\begin{eqnarray}
\langle {\cal O}({\mathrm x}_s ) \rangle ={{\bf  E}_t\Big ({\cal O}({\mathrm x}_s )\ e^{\int_0^t u({\mathrm x}_s) ds} \Big )\over 
{\bf E}_t\Big (e^{\int_0^t u({\mathrm x}_s) ds} \Big )}, 
~~~~0 \leq s \leq t \label{stochI}
\end{eqnarray}
where we define via (\ref{B}), (\ref{measure1})
\begin{eqnarray}
&& {\bf E}_t\Big ({\cal O}({\mathrm x}_s)\Big )=\int  d{\bf w_t}\ d{\bf M}\ {\cal O}({\mathrm x}_s) ~~~~0 \leq s \leq t . \label{stochII} 
\end{eqnarray}
(\ref{stochI}) can be used provided that solutions of the associated SDEs 
are available, so that the fields ${\mathrm x}_{tj}$  are expressed in terms of the variables ${\mathrm w}_{tj}$.

\section{The quantum (D)NLS and a hierarchy of S(P)DEs}

\noindent 
We start our analysis with the DNLS model, with the corresponding quantum Lax operator given by \cite{KunduRagnisco},  \cite{Sklyanin},
\begin{eqnarray}
 L_j(\lambda) = \begin{pmatrix}
\lambda + \Theta_j + z_jZ_j     & z_j \nonumber \\
 Z_j                              & 1  \end{pmatrix}
\end{eqnarray}
$z_j,\  Z_j$ are canonical  $[z_i,\ Z_j]= -\delta_{ij}$, and we consider the map:
\begin{equation}
z_ j\mapsto {\mathrm x}_j,~~~~~Z_j \mapsto \partial_{{\mathrm x}_j}. \label{map1}
\end{equation}

\noindent Let us now define the generating function of the integrals of motion of the system:
\begin{equation}
{\mathbf t}(\lambda) = tr\Big ( L_M(\lambda)\ldots L_2(\lambda) L_1(\lambda)\Big ).
\end{equation}
Indeed, the expansion of $\ln {{\bf t}(\lambda)} = \sum_{k=0}^M {I_k \over \lambda^k}$ provides the local integrals of motion (see e.g. \cite{Korepin}).
We keep here terms up to third order in the expansion of $ \ln{{\bf t}}$ and  by suitably scaling the involved fields, 
we obtain the first three local integrals of motion of the quantum DNLS hierarchy (keeping the suitably scaled terms) \cite{KunduRagnisco}:
\begin{eqnarray}
H_1 &=& \sum_j^M {\mathrm x}_j \partial_{{\mathrm x}_j} \nonumber\\
H_2 &=&  {1\over 2} \sum_{j=1}^M {\mathrm x}_j^2 \partial_{{\mathrm x}_j}^2 - \sum_{j=1}^M 
\Delta^{(1)}({\mathrm x}_j)\partial_{{\mathrm x}_j}   \nonumber\\
H_3 &=&  {1\over 2} \sum_{j=1}^M {\mathrm x}_j^2 \partial_{{\mathrm x}_j}^2 - 
\nu \sum_{j=1}^M \Delta^{(2)}({\mathrm x}_j)\partial_{{\mathrm x}_j}  + \mbox{(higher order terms)}...
\label{eq:ALP_ExDNLSP} 
\end{eqnarray}
where we have chosen $\Theta_j = 1$, and 
$H_1 =I_{1},\ H_2 =-I_2 +{1\over 2} I_1,\ H_3= - {1\over 3} \big (I_3 +I_2 - {1\over 2} I_{1})$, $\nu = {1\over 3}$. We also define:
$ \Delta^{(1)}z_j=z_{j+1} -z_j,  ~~ \Delta^{(2)}z_j= z_{j+2} -2z_{j+1} + z_j$.\\
The next order in the expansion provides $H_4$, which  is the Hamiltonian of the quantum version of  complex mKdV system and so on. 
The equations of motion (classical and quantum) associated e.g. to $H^{(2)}$ can be derived via the zero curvature 
condition or Heisenberg's equation (recall also (\ref{map1})):
\begin{equation}
{dz_j \over dt}  = -  \Delta^{(1)}z_j + z^2_jZ_j.\label{qem}
\end{equation}
Similar equations can be obtained for $H_3$, but are omitted here for brevity.
The Hamiltonians $H_{2,3}$ are of the form (\ref{general}), and the corresponding set of SDEs are \cite{DoikouMalhamWiese}
\begin{equation}
d{\mathrm x}_{tj}  =- \nu_k \Delta^{(k-1)}{\mathrm x}_t dt + {\mathrm x}_{tj} d{\mathrm w}_{tj}. \label{SDE1}
\end{equation}
where $k \in \{ 2,\ 3\}$ and $\nu_2 =1,\ \nu_3 = {1\over 3}$.  $\nu_k$ can be set equal to one henceforth, after suitably rescaling  time. 
By comparing (\ref{qem}) and (\ref{SDE1})  ($k=2$) we observe that the non-linearity 
appearing in (\ref{qem})  is replaced by the multiplicative noise in (\ref{SDE1}).

Let us now derive the solution of the set of SDEs (\ref{SDE1})  introducing suitable integrator factors (see e.g. \cite{Oksendal}). 
Let us consider the general set of SDES
\begin{equation}
d{\mathrm x}_{tj} =b_j({\mathrm x}_t) dt + {\mathrm x}_{tj} d{\mathrm w}_{tj}. \nonumber
\end{equation}
We introduce the following set of integrator factors:
\begin{equation}
{\cal F}_j(t) = \exp\Big ( -\int_0^t d{\mathrm w}_{sj} + {1\over 2}\int_0^t ds \Big ) \label{change} 
\end{equation}
and define the new fields: ${\mathrm y}_{tj} = {\cal F}_{j}(t) {\mathrm x}_{tj}$, then one obtains 
a differential equation for the vector field ${\mathrm y}$:
\begin{equation}
{d{\mathrm y}_t\over dt}= {\cal A}(t) {\mathrm y}_t\ \Rightarrow {\mathrm y}_t = 
{\cal P}\exp\Big ( \int_{0}^t {\cal A}(s) ds \Big ) {\mathrm y}_{0}.\label{lode}
\end{equation}
For instance in the case of  (\ref{SDE1}), for $k=2$, the $M \times M$  matrix ${\cal A}$ is given as
\begin{equation}
{\cal A}(t) = \sum_{j=1}^M \Big ( e_{jj} - {\cal B}_j(t)e_{j j+1}\Big), ~~~~~~~
{\cal B}_j(t) = \exp \Big (\Delta^{(1)}({\mathrm w}_{tj}) \Big ),  \nonumber
\end{equation} 
where $e_{ij}$ are $M\times M$ matrices with entries $(e_{ij})_{kl} = \delta_{ik} \delta_{jl}$. For $k=3$,
the ${\cal A}$ matrix  involves also terms $e_{jj+2}$, and so on.
The solution (\ref{lode})  can be expressed as a formal series expansion
\begin{eqnarray}
&&{\cal P}\exp\Big ( \int_{0}^t {\cal A}(s) ds \Big ) =  \nonumber\\ &&\sum_{n=0}^{\infty} \int_0^t \int_0^{t_n} \ldots 
\int_{0}^{t_2} d{t_n} dt_{n-1}  \ldots dt_{1} {\cal A}(t_n) {\cal  A}(t_{n-1}) \ldots {\cal  A}(t_1),  \nonumber\\ 
&& t\geq t_{n} \geq t_{n-1} \ldots \geq t_{2}.  \nonumber
\end{eqnarray}

\noindent {\bf Remark 1.}\\
The discrete version of the stochastic Burgers equation can be obtained from the discrete stochastic heat equation through 
the analogue of the Cole-Hopf transformation. Indeed,  by setting
${\mathrm x}_j = e^{{\mathrm y}_j}$,  in (\ref{SDE1}) ($k=3$):
\begin{equation}
d{\mathrm y}_j = - \Big ( e^{\Delta{\mathrm y}_j}\big (  e^{\Delta{\mathrm y}_{j+1}} -1\big ) -
 \big (e^{\Delta{\mathrm y}_{j}} +1\big ) \Big ) dt+ d{\mathrm w}_j,  \label{HJ}
\end{equation}
where for simplicity we have set $\Delta^{(1)} = \Delta$.
By also setting  $u_j = \Delta {\mathrm y}_{j}$, we obtain a discrete version of the stochastic Burgers equation
\begin{equation}
d u_j =-\Big (  e^{u_{n+1}} \big ( e^{u_{j+2}}-  e^{u_{j}} \big ) -2  \big ( e^{u_{j+1}}-  e^{u_{j}} \big ) 
\Big ) dt+ \Delta d{\mathrm w}_j. \label{Burgers}
\end{equation}
Assuming the scaling $\Delta {y_j} \sim \delta$, we expand the exponentials and keep up to second order terms in (\ref{HJ}), (\ref{Burgers}):
\begin{eqnarray}
&&d{\mathrm y}_j = -\Big (\Delta^{(2)} {\mathrm y}_j + \big (\Delta {\mathrm y}_j \big)^2 + {\cal O}(\delta^3 )\Big ) +  d{\mathrm w}_j\\
&& d u_j =  -\Big (\Delta^{(2)} u_j + \Delta u_j^2 + {\cal O}(\delta^3) \Big )+ \Delta d{\mathrm w}_j.
\end{eqnarray}
The second of the equations above provides a good approximation for the discrete viscous  Burgers equation, 
as will be also clear in the next subsection.

\subsection{The continuum models and SPDEs}

\noindent It will be instructive to consider the continuum limits of the Hamiltonians  $H_2,\ H_3$ 
(\ref{eq:ALP_ExDNLSP}) and the respective SDEs.  After considering the thermodynamic limit 
$M \to \infty,\  \delta \to 0$ $(\delta \sim {1\over M}$) we obtain
\begin{eqnarray}
 && {\mathrm x}_{tj}\  \to\   \varphi(x, t), ~~~~ {{\mathrm x}_{tj+1} - {\mathrm x}_{tj} \over \delta}\  \to\  \partial_x \varphi(x,t) ,\nonumber\\ 
&&\delta \sum_j  f_j\  \to\  \int dx\ f(x),~~~~~{\mathrm w}_{tj}\  \to\  W(x, t), \label{dictionary}
\end{eqnarray}
where the Wiener field or Brownian sheet $W(x,t)$ is periodic and square integrable in $[-L,\  L]$, and is represented as \cite{Prevot}
\begin{equation}
W(x,t) = {\sqrt{L} \over \pi}\sum_{n\geq 1}{1\over n} \Biggl ( X_t^{(n)} \cos{n\pi x \over L}+ Y_t^{(n)} \sin{n\pi x \over L}\Biggr ), \label{sheet1} 
\end{equation}
$X_t^{(n)},\  Y_t^{(n)}$ are independent Brownian motions.
In the continuum limit the Hamiltonians  (\ref{eq:ALP_ExDNLSP}) become the Hamiltonians of quantum NLS hierarchy:
\begin{eqnarray}
&&H_c^{(k)}= \int dx\ \Big ( {1\over 2} \varphi^2(x) \hat \varphi^2(x) - \partial^{(k-1)}_x \varphi(x) \hat \varphi(x)\Big ),  ~~~k = 2,\ 3
\end{eqnarray}
where $\big [\varphi(x),\ \hat \varphi(y)\big ] = \delta(x-y)$, 
($\hat \varphi(x) \sim {\partial\over \partial \varphi(x)}$) 
and the SDEs (\ref{SDE1}) become the stochastic transport ($k=2$) and heat equation ($k=3$) with multiplicative noise:
\begin{equation}
\partial_t \varphi(x,t) = -\partial^{k-1}_x \varphi(x, t) + \varphi(x,t)\dot{W}(x,t). \label{heats}\nonumber
\end{equation}
The stochastic heat equation  can be mapped to the stochastic Hamilton-Jacobi and viscous Burgers equations \cite{Corwin}. 
Indeed, we set: $\varphi = e^h ,\  u = \partial_x h $ then (\ref{heats}): 
\begin{eqnarray}
&&  \partial_t  h(x,t) = - \partial^2_x h(x, t) - (\partial_x h(x,t))^2+  \dot{W}(x,t)\nonumber\\
&& \partial_t  u(x,t) = - \partial^2_x u(x, t) -2u(x,t) \partial_x u(x,t)+\partial_x \dot{W}(x,t).
\end{eqnarray}
Connections between the SDEs and the quantum Darboux transforms \cite{Korff}, \cite{DoikouFindlay} can be also studied.
The classical Darboux-B\"{a}cklund transformation \cite{ZakharovShabat}, \cite{MatveevSalle} provides an efficient way to 
find solutions of integrable PDEs.The key question is how this transformation can facilitate the solution of 
SDEs \cite{Corwin}, \cite{oconnell}, \cite{DMSW}.\\

\noindent {\bf Acknowledgments}\\
AD acknowledges support from the  EPSRC research grant:  EP/R009465/1.

%
%
\footnotesize{
}
\end{document}